# Saddle-point-Saddle-focus singular cycles and existence of horseshoes

Xiao-Song Yang

School of Mathematics and Statistics,,
Huazhong University of Science and Technology.
Wuhan, 430074, China
yangxs@mail.hust.edu.cn

*Abstract*

In this paper we study a type of two singular point singular cycle where one heteroclinic orbit is the transversal intersection of the 2-dimensional stable manifold of one singular point and the 2-dimensional unstable manifold of other singular point. We show that this kind of singular cycle can give rise to a new mechanism of creation of chaos in 3-dimensional vector fields.

## 1 Introduction

It is well known that singular cycles are a remarkable mechanism for the creation of chaos in dynamical systems [1-3]. One of the famous singular cycles is the homoclinic cycle, by which the Shilnikov phenomena [4,5] can take place. Another type of singular cycles is the heteroclinic cycle, which gives rise to another kind of mechanism of creation of chaos [7].

Recall that a cycle of a smooth vector field $F$ is a compact invariant chain recurrent set of $F$ consisting of a finite family of critical elements and orbits whose $\omega$-limit set and $\alpha$-limit set are critical elements of this family. Here by critical elements we mean periodic orbits or singular points.

In this paper we are only concerned with a special class of cycles, the singular cycles.

**Definition 1** A singular cycle of a vector field $F$ is a compact invariant chain recurrent set of $X$ consisting of finite number of singular points and the orbits whose $\omega$-limit set and $\alpha$-limit set are different points among these singular points.

If the singular point is unique, then the singular cycle is the so called homoclinic cycle or homoclinic orbit [7]. In case the number of singular points is not less than two, we have a heteroclinic cycle [7].

The aim of this paper is to study how a singular cycle with only two singular points providing creation of chaos. For this class of singular cycles, there have some results on chaos created by the singular cycle where every heteroclinic orbit is contained in a 1-dimensional stable (unstable) manifold of one of the singular points [6]. In this paper we consider a new type of two singular point singular cycle where one heteroclinic orbit is the transversal intersection of the 2-dimensional stable manifold of one singular point and the 2-dimensional unstable manifold of other singular point. We will show that this kind of singular cycle give rise to another mechanism of creation of chaos in 3-dimensional vector fields.

## 2 Some preliminaries

In this section we recall some basic elements in chaos theory of dynamical systems.

Let $M$ and $N$ be topological spaces and consider two continuous maps $f: M \to M$ and $g: N \to N$. The map $f$ is said to be semi-conjugate to $g$, if there is a continuous surjective map $h: M \to N$ such that

$$h \circ f = g \circ h \qquad (2.1)$$

If $g: N \to N$ is replaced by $m$-shift $\sigma_m : \Sigma_m \to \Sigma_m$ and (2.1) holds true for the $m$-shift, then $f$ is said to have a topological horseshoe.

An important fact about the semi-conjugacy is that the topological entropy of $f$ is not less than that of $g$. In particular if $g$ is chaotic in terms of positive entropy then $f$ is chaotic.

For convenience of main arguments of this paper we first recall an elementary fact.
Let $A \subset R^2$ be a rectangular as shown with $A^l$ and $A^r$ being its left side and right side, respectively, and $A^t$ and $A^b$ being its top side and bottom side, respectively:

$$A = \{w = (x, y) \in R^2 : a \leq x \leq b, c \leq y \leq d\}$$

Suppose $S_i$, $i = 1, ..., m$, is a quadrilateral of $A$ with its left and right side contained in $A^l$ and $A^r$, respectively and its top and bottom side contained in the interior of $A$. Consider a piece wise continuous map

$$f(w) = (f_1(w), f_2(w)) : A \to R^2.$$

We have the following fact.

**Lemma** Suppose that for every $1 \leq i \leq m$, $f$ restricted to every $S_i$ is continuous, and satisfies

$$a < f_1(w) < b, \ w \in S_i, \ i = 1, ..., m$$

$$f_2(w) < c, \ w \in S_i^b \quad f_2(w) > d, \ w \in S_i^t$$

or

$$f_2(w) > c, \ w \in S_i^b \quad f_2(w) < d, \ q \in S_i^t$$

Then there exists a compact invariant set $\Lambda \subset A$ such that $P|\Lambda$ is semi-conjugate to the $m$-shift dynamics.

This statement is very easy to prove [7], it is also a consequence of The topological horseshoe lemma given in [8].

## 3 Main results

Consider the following dynamical system

$$\dot{x} = F(x), \quad x \in R^3 \qquad (1)$$

where $F(x)$ is a continuous differentiable vector field on $R^3$. Suppose (1) has two singular points $p$ and $q$ that satisfy the following assumptions.

**A1** At the point $p$, the linearized flow of (1) in an appropriate local coordinate system is given by the following equations

$$\begin{cases} \dot{x}_p = \lambda x_p \\ \dot{y}_p = \lambda y_p \\ \dot{z}_p = -\tilde{\lambda} z_p \end{cases} \quad \lambda > 0 \text{ and } \tilde{\lambda} > 0 \qquad (2)$$

**A2** At the point $q$ the linearized flow of (1) in an appropriate local coordinate system is given by the following equations

$$\begin{pmatrix} \dot{x}_q \\ \dot{y}_q \\ \dot{z}_q \end{pmatrix} = \begin{bmatrix} -\alpha & -\beta & 0 \\ \beta & -\alpha & 0 \\ 0 & 0 & \rho \end{bmatrix} \begin{pmatrix} x_q \\ y_q \\ z_q \end{pmatrix} \quad \alpha > 0 \quad \rho > 0 \qquad (3)$$

**A3** The 2-dimensional unstable manifold of $p$ intersects transversely with the 2-dimensional stable manifold of $q$. The intersection is a heteroclinic orbit and denoted by $\psi(t)$.

**A4** There exists a heteroclinic orbit $H(t)$ of (1) from $q$ to $p$. Furthermore assume that locally $H(t) = (0,0,z_q(t)), z_q(t) > 0$ in a neighborhood of $q$, which is contained in $W_{loc}^u(q)$, and $H(t) = (0,0,z_p(t)), z_p(t) > 0$ in a neighborhood of $p$, which is contained in $W_{loc}^s(p)$.

With the assumptions given above, (1) has a singular cycle $\Theta = p \cup q \cup \psi(t) \cup H(t)$

In a neighborhood of the point $p$, we define two small cross sections to the flow of (1) in terms of the appropriately chosen local coordinate system:

$$\pi_1 = \{(x_p, y_p, z_p) : x_p^2 + y_p^2 = \bar{r}_p^2, 0 < z_p < \delta_p\}$$

$$\pi_2 = \{(x_p, y_p, z_p) : x_p^2 + y_p^2 < \bar{r}_p^2, z_p = \delta_p\}$$

such that the section $\pi_2$ intersects transversely with $H(t)$ at unique point. This can be the case if $\bar{r}_p$ and $\delta_p$ small enough. Without loss of generality it is assumed that the intersection point is $(x_p, y_p, z_p) = (0, 0, \delta_p)$ in the local coordinate system. In view of (2), the unstable manifold of $p$ can be locally expressed as

$$W_{loc}^u(p) = \{(x_p, y_p, z_p) : z_p = 0\}.$$

In a neighborhood of the point $q$, define two small cross sections to the flow of (1) in terms of the appropriately chosen local coordinate system:

$$\pi_3 = (x_q, y_q, z_q) : x_q^2 + y_q^2 < \bar{r}_q^2, z_q = \delta_q$$

$$\pi_4 = \{(x_q, y_q, z_q) : x_q^2 + y_q^2 = \bar{r}_q^2, 0 < z_q < \delta_q\}$$

such that the section $\pi_3$ intersects transversely with $H(t)$ at unique point. This can be the case if $\bar{r}_q$ and $\delta_q$ are small enough. Without loss of generality it is assumed that the intersection point is $(x_q, y_q, z_q) = (0, 0, \delta_q)$ in the local coordinate system. In view of (3), the stable manifold of $q$ can be locally expressed as

$$W_{loc}^s(q) = \{(x_q, y_q, z_q) : z_q = 0\}$$

Now we consider the map $P_1 : \pi_2 \to \pi_1$ induced by the flow of (1). Since the flow of (1) in a neighborhood of $p$ can be described by (2) in the following form

$$\begin{cases} x_p(t) = x_p(0)e^{\lambda t} \\ \dot{y}_p(t) = y_p(0)e^{\lambda t} \\ \dot{z}_p(t) = z_p(0)e^{-\tilde{\lambda} t} \end{cases} \quad (4)$$

the map $P_1 : \pi_2 \to \pi_1$ can be defined as follows.

For each point $\xi = (x_p, y_p, \delta_p) \in \pi_2$, the time $T$ to send $\xi$ to $\pi_1$ by the flow (4) satisfies (where we regard $\xi$ as initial point):

$$(x_p^2 + y_p^2)e^{2\lambda T} = \bar{r}_p^2, \quad T = \frac{1}{2\lambda} \ln \frac{\bar{r}_p^2}{x_p^2 + y_p^2}$$

Therefore

$$P_1(\xi) = (x_p e^{\lambda T}, y_p e^{\lambda T}, \delta_p e^{-\tilde{\lambda} T})^T$$

$$= (\frac{x_p \bar{r}_p}{\sqrt{x_p^2 + y_p^2}}, \frac{y_p \bar{r}_p}{\sqrt{x_p^2 + y_p^2}}, \delta_p \left( \frac{\sqrt{x_p^2 + y_p^2}}{\bar{r}_p} \right)^{\tilde{\lambda}/\lambda} )^T. \quad (5)$$

In a neighborhood of $q$, the flow of (1) is governed by (3) and can be given by

$$\begin{cases} x_q(t) = e^{-\alpha t}(x_q(0) \cos \beta t - y_q(0) \sin \beta t) \\ \dot{y}_q(t) = e^{-\alpha t}(x_q(0) \sin \beta t + y_q(0) \cos \beta t) \\ \dot{z}_q(t) = z_q(0) e^{\rho t} \end{cases} \quad (6)$$

Thus the map $P_3 : \pi_4 \to \pi_3$ can be defined as follows.

For each point $(x_q, y_q, z_q) \in \pi_4$, the time $T$ to send it to $\pi_3$ by the flow of (6) satisfies

$$z_q e^{\rho T} = \delta_q, \quad T = \frac{1}{\rho} \ln \frac{\delta_q}{z_q}$$

The map can then be written as

$$P_3(x_q, y_q, z_q) = ( \left(\frac{z_q}{\delta_q}\right)^{\alpha/\rho} (x_q \cos \vartheta - y_q \sin \vartheta), \left(\frac{z_q}{\delta_q}\right)^{\alpha/\rho} (x_q \sin \vartheta + y_q \cos \vartheta), \delta_q )^T$$

where $\vartheta = \vartheta(z_q) = \frac{\beta}{\rho} \ln \frac{\delta_q}{z_q}$,

Setting $x_q = \bar{r}_q \cos \theta$ and $y_q = \bar{r}_q \sin \theta$ we have

$$P_3(\theta, z_q) = (\bar{r}_q \left(\frac{z_q}{\delta_q}\right)^{\alpha/\rho} \cos(\theta + \vartheta), \bar{r}_q \left(\frac{z_q}{\delta_q}\right)^{\alpha/\rho} \sin(\theta + \vartheta), \delta_q )^T$$

And in $(\theta, r_q)$-coordinates the map can also be written as

$$P_3(\theta, z_q) = (\theta + \vartheta, \bar{r}_q \left(\frac{z_q}{\delta_q}\right)^{\alpha/\rho}, \delta_q )^T \quad (7)$$

Furthermore, in the $(\theta, z_q)$-coordinates for the cross section $\pi_4$ it can be seen that there is a fixed $\theta_q$ such that $(\theta, z_q) = (\theta_q, 0)$ corresponds to the point $q_\psi$ at which the heteroclinic orbit $\psi(t)$ intersects with $\pi_4$.

For convenience of the later discussions, we need to consider a cross subsection $\Pi_q \subset \pi_4$

defined as follows:

$$\Pi_q = \{(\theta, z_q) : 0 \leq |\theta - \theta_p| \leq \eta, 0 < z_q < \delta_q\} \qquad (8)$$

where $\eta$ is a sufficiently small positive number. The cross section $\Pi_q$ is illustrated in Fig.1

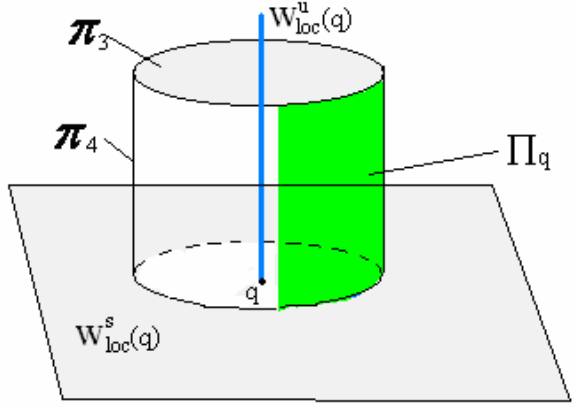

Fig.1 The cross section $\Pi_q$

**Remark 1** Geometrically the image $P_3(\Pi_q)$ is a logarithmic spiral-like band in $\pi_3$, and the number of times this image turns around $r_q = 0$ goes to infinity as the variable $z_q$ approaches to zero, as shown in Fig.2.

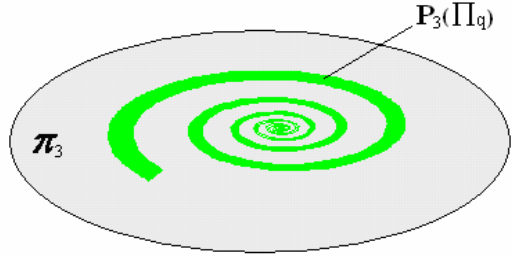

Fig.2 $P_3(\Pi_q)$ is a logarithmic spiral like band in $\pi_3$,

Now consider the map $P_2 : \pi_3 \to \pi_2$. This map is induced by the flow of (1) in the vicinity of heteroclinic orbit $H(t)$ from $q$ to $p$ and can be approximated by the following linear map (still denote it by $P_2$ for simplicity)

$$P_2(x_q, y_q, \delta_q) = (ax_q + by_q, cx_q + dy_q, \delta_p)^T. \qquad (9)$$

where $a$, $b$, $c$, and $d$ are constants and satisfy $ad - bc \neq 0$. Note that the orientation of $\pi_3$ is reversed under the map $P_2$ in the given local coordinates. Thus the map $P_2 \circ P_3 : \Pi_q \to \pi_2$ can be illustrated in Fig.3

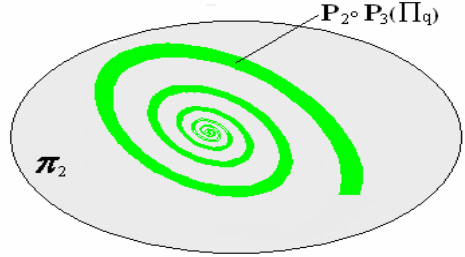

Fig.3 The image of $\Pi_q$ under the map $P_2 \circ P_3 : \Pi_q \to \pi_2$

The map $P_4 : \pi_1 \to \pi_4$ induced from the flow of (1) is generally not well defined. However we can take a piece of $\pi_1$ to define $P_4$. Since the intersection of unstable manifold of $p$ with the stable manifold of $q$ is the heteroclinic orbit $\psi(t)$ from $p$ to $q$, we will take the piece attached to this orbit.

Without loss of generality suppose that in the local coordinate system around $p$ and the local coordinate system around $q$, the unstable manifold of $p$ and the stable manifold of $q$ intersect with each other perpendicularly.

Now let $x_p = \bar{r}_p \cos\phi$, $y_p = \bar{r}_p \sin\phi$, then in the $(\phi, z_p)$ – coordinate system we have

$$\pi_1 = \{(\phi, z_p) : 0 \le \phi \le 2\pi, 0 < z_p < \delta_p\}.$$

**Remark 2** It is easy to see that there is a fixed $\phi_p$ such that $(\phi, z_p) = (\phi_p, 0)$ corresponds to the point $p_\psi$ at which the heteroclinic orbit $\psi(t)$ intersects with $\pi_1$.

Take a small number $\nu > 0$, consider the piece of the cross section $\Pi_p \subset \pi_1$ defined as follows

$$\Pi_p = \{(\phi, z_p) : 0 \le |\phi - \phi_p| \le \nu, 0 < z_p < \delta_p\} \qquad (10)$$

The linearization of the map $P_4 : \Pi_p \to \pi_4$ can be expressed as (still indicated by $P_4$)

$$P_4(\phi, z_p) = (hz_p + \theta_q, g(\phi - \phi_p)) \qquad (11)$$

i.e. $(\theta, z_q) = (hz_p + \theta_q, g(\phi - \phi_p))$, where $h$ and $g$ are constants and satisfy $hg < 0$

**Remark 3** It is easy to see from **Remark 1** and **Remark 2** that

$$P_4(\phi_p, 0) = (\theta_q, 0), \text{ i.e., } P_4(p_\psi) = q_\psi. \tag{12}$$

In view of the above discussions, geometrically there are two possible situations happened to the map $P_4 : \Pi_p \to \pi_4$: the way the image $P_4(\Pi_p)$ located in $\pi_4$ can be illustrated in Fig.4, where the line $l$ indicates the intersection of $P_4(\Pi_p)$ with the stable manifold of $q$.

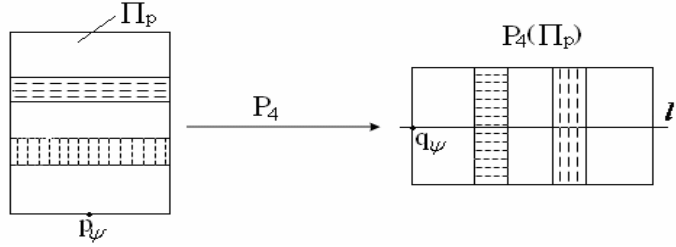

a. Situation 1

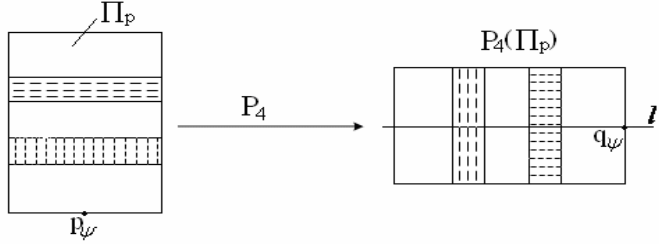

b Situation 2

Fig.4 There are two possible situations for the map $P_4$

Now we can have the following statement.

**Theorem 1** For each positive integer $n$, there exists an invariant compact set in $\Lambda_n \subset \Pi_q$ such that $P : \Lambda_n \to \Lambda_n$ is conjugate with $n$-shift symbolic dynamics provided the assumptions **A1**-**A4** are satisfied, where $P : \Pi_q \to \Pi_q$ is defined as follows

$$P = P_4 \circ P_1 \circ P_2 \circ P_3.$$

*Proof* Note that we can take the number $\nu > 0$ and $\delta_p$ small enough, so that on the cross section $\Pi_p \subset \pi_1$ defined by (9) the map $P_4$ can be characterized by (11).

Now consider the map $P_3$. For any positive integer $n$ and a number $\varsigma > 0$, it can be seen that there exist two numbers $0 < \bar{\varsigma}_n < \varsigma_n < \delta_q$ such that

$$\vartheta(\bar{\varsigma}_n) - \vartheta(\varsigma_n) = \frac{\beta}{\rho} \ln \frac{\varsigma_n}{\bar{\varsigma}_n} \geq 2n\pi + 2\eta \quad \text{and} \quad \bar{r}_q \left( \frac{\varsigma_n}{\delta_q} \right)^{\alpha/\rho} < \varsigma \qquad (13)$$

This implies for each radial line emanating form $r_q = 0$ in the cross section $\pi_3$, its intersection with the image $P_3(\Pi_\varsigma(\bar{\varsigma}_n, \varsigma_n))$ consists of at least $n$ connected components, where

$$\Pi_\varsigma(\bar{\varsigma}_n, \varsigma_n) = \{(\theta, z_q) : 0 \leq |\theta - \theta_q| \leq \eta, \bar{\varsigma}_n < z_q < \varsigma_n\}. \qquad (14)$$

In addition, the following statement holds for the cross section defined in (14).

**Fact 1** For any $\bar{\varepsilon} > 0$, there exists a number $\varepsilon(\bar{\varepsilon}) > 0$, such that for every point

$$(\phi, z_p) \in \Pi_p \cap P_1 \circ P_2 \circ P_3(\Pi_\varsigma(\bar{\varsigma}_n, \varsigma_n))$$

there holds $z_p < \bar{\varepsilon}$, provided that the parameter $\varsigma$ defining $\Pi_\varsigma(\bar{\varsigma}_n, \varsigma_n)$ satisfies $\varsigma < \varepsilon(\bar{\varepsilon})$.

This can be seen from the composition map $\bar{P} = P_1 \circ P_2 \circ P_3$, in fact, we have

$$P_2 \circ P_3(\theta, z_q) = \begin{pmatrix} \bar{r}_q \left( \frac{z_q}{\delta_q} \right)^{\alpha/\rho} (a\cos(\theta + \vartheta) + b\sin(\theta + \vartheta)), \\ \bar{r}_q \left( \frac{z_q}{\delta_q} \right)^{\alpha/\rho} (c\cos(\theta + \vartheta) + d\sin(\theta + \vartheta)), \\ \delta_p \end{pmatrix}$$

Let $k = \bar{r}_q \left( \frac{z_q}{\delta_q} \right)^{\alpha/\rho}$, then

$$\bar{P} = P_1 \circ P_2 \circ P_3$$

$$= \begin{pmatrix} \frac{k(a\cos(\theta + \vartheta) + b\sin(\theta + \vartheta))\bar{r}_p}{k\sqrt{(a\cos(\theta + \vartheta) + b\sin(\theta + \vartheta))^2 + (c\cos(\theta + \vartheta) + d\sin(\theta + \vartheta))^2}}, \\ \frac{k(c\cos(\theta + \vartheta) + d\sin(\theta + \vartheta))\bar{r}_p}{k\sqrt{(a\cos(\theta + \vartheta) + b\sin(\theta + \vartheta))^2 + (c\cos(\theta + \vartheta) + d\sin(\theta + \vartheta))^2}}, \\ \delta_p \left( \frac{k\sqrt{(a\cos(\theta + \vartheta) + b\sin(\theta + \vartheta))^2 + (c\cos(\theta + \vartheta) + d\sin(\theta + \vartheta))^2}}{\bar{r}_p} \right)^{\tilde{\lambda}/\lambda} \end{pmatrix}$$

In particular, the third component of the above vector is

$$\delta_p\left(\frac{z_q}{\delta_q}\right)^{\frac{\alpha\tilde{\lambda}}{\rho\lambda}}\left((a\cos(\theta+\vartheta)+b\sin(\theta+\vartheta))^2+(c\cos(\theta+\vartheta)+d\sin(\theta+\vartheta))^2\right)^{\tilde{\lambda}/2\lambda}$$

which approaches zero as $z_p \to 0$.

In what follows we will be showing existence of disjoint compact subsets of $\Pi_q$ to guarantee existence of topological horseshoe. First, note that image $\overline{P}(\Pi_q) = P_1 \circ P_2 \circ P_3(\Pi_q)$ goes around in $\pi_1$ in a manner as illustrated as in Fig.5

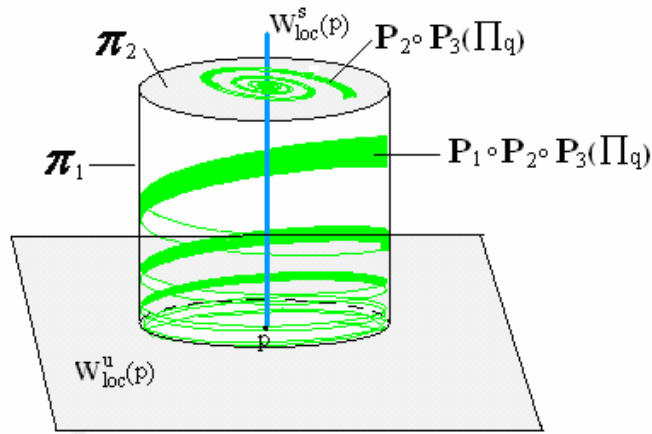

Fig.5 the way the image $\overline{P}(\Pi_q)$ located in $\pi_1$.

It is now can be see that the image

$$\Pi_p \cap P_1 \circ P_2 \circ P_3(\Pi_\varsigma(\overline{\varsigma}_n,\varsigma_n)) \quad \Pi_p \cap \overline{P}(\Pi_\varsigma(\overline{\varsigma}_n,\varsigma_n)) \quad (15)$$

consists of at least $n$ disjoint quadrilateral $S_i$, $i=1,2,...,n$, as illustrated in Fig.6

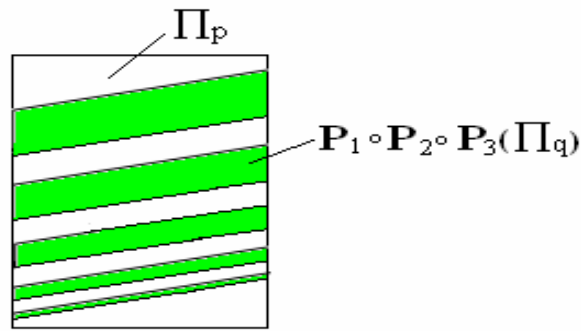

Fig.6 The quadrilaterals $S_i$ in $\Pi_p$

Now denote the "left" and "right" sides of $\Pi_p$ as follows

$$L_p^l = \{(\phi, z_p) : \phi = \phi_p - \nu, 0 < z_p < \delta_p\} \tag{16}$$

$$L_p^r = \{(\phi, z_p) : \phi = \phi_p + \nu, 0 < z_p < \delta_p\} \tag{17}$$

Then clearly the "left" and "right" sides of quadrilateral $S_i$, are of the following form

$$S_i^l = S_i \bigcap L_p^l \quad S_i^r = S_i \bigcap L_p^r,$$

and the "top" and "bottom" sides, denoted by $S_i^t$ and $S_i^b$ are clearly contained in $\Pi_p$.

Let $\overline{P} = P_1 \circ P_2 \circ P_3$, which is clearly a diffeomorphism from $\Pi_q$ to its image. It is apparent that

$$\overline{S}_i = \overline{P}^{-1}(S_i) \subset \Pi_\varsigma(\overline{\varsigma}_n, \varsigma_n) \tag{18}$$

and these four sets $\overline{P}^{-1}(S_i^l)$, $\overline{P}^{-1}(S_i^r)$, $\overline{P}^{-1}(S_i^t)$ and $\overline{P}^{-1}(S_i^b)$ are obviously contained in $\Pi_\varsigma(\overline{\varsigma}_n, \varsigma_n)$, and are four sides of $\overline{S}_i$. Consider the two sides of $\Pi_\varsigma(\overline{\varsigma}_n, \varsigma_n)$:

$$L_q^l = \{(\theta, z_q) : \theta = \theta_q - \eta, \overline{\varsigma}_n < z_q < \varsigma_n\}$$

$$L_q^r = \{(\theta, z_q) : \theta = \theta_q + \eta, \overline{\varsigma}_n < z_q < \varsigma_n\}$$

It is easy to see that

$$\overline{P}^{-1}(S_i^t) \subset L_q^l \text{ and } \overline{P}^{-1}(S_i^b) \subset L_q^r \tag{19}$$

or

$$\overline{P}^{-1}(S_i^t) \subset L_q^r \text{ and } \overline{P}^{-1}(S_i^b) \subset L_q^l \tag{20}$$

depending on the sign of $\beta$ in (6). Without loss of generality, we consider the case (19).

In this case let

$$\overline{S}_i^l = \overline{P}^{-1}(S_i^t) \text{ and } \overline{S}_i^r = \overline{P}^{-1}(S_i^b) \tag{21}$$

and let

$$\overline{S}_i^t = \overline{P}^{-1}(S_i^l), \quad \overline{S}_i^b = \overline{P}^{-1}(S_i^r). \tag{22}$$

It is obvious that $\overline{S}_i^t \subset \Pi_\varsigma(\overline{\varsigma}_n, \varsigma_n)$ and $\overline{S}_i^b \subset \Pi_\varsigma(\overline{\varsigma}_n, \varsigma_n)$.

In view of the formula $(\theta, z_q) = P_4(\phi, z_p) = (hz_p + \theta_q, g(\phi - \phi_p))$ in (10) the following fact is obvious.

**Fact 2** There exists an $0 < \overline{\varepsilon}_p < \delta_p$ and $0 < \overline{\varepsilon}_q \leq \delta_q$, such that for every point $(\phi, z_p) \in \Pi_p$ satisfying $0 < z_p < \bar{\varepsilon}_p$, there holds $|\theta - \theta_q| < \eta$, and for $(\phi, z_p) \in L_p^l$, there holds $z_q \geq \bar{\varepsilon}_q$ if $g > 0$ or for $(\phi, z_p) \in L_p^r$, $z_q \geq \bar{\varepsilon}_q$ if $g < 0$, where

$$\theta = h z_p + \theta_q \quad \text{and} \quad z_q = g(\phi - \phi_p) \qquad \bar{\varepsilon}_q$$

Keeping in mind **fact 1** and $\varepsilon_p$ in **fact 2,** take $\varsigma$ satisfying $\varsigma < \varepsilon(\bar{\varepsilon}_p)$ and $\varsigma < \bar{\varepsilon}_q$ and consider the corresponding cross section $\Pi_\varsigma(\bar{\varsigma}_n, \varsigma_n) \subset \Pi_q$. It is easy to see that the compact sets $\bar{S}_i \subset \Pi_\varsigma(\bar{\varsigma}_n, \varsigma_n), i = 1,2,\ldots,n$ defined in (18) are disjoint. Furthermore, for each $\bar{S}_i$

$$\bar{P}(\bar{S}_i) = P_1 \circ P_2 \circ P_3(\bar{S}_i) = S_i.$$

Recall that $\bar{S}_i^l = \bar{P}^{-1}(S_i^t) \subset L_q^l$ and $\bar{S}_i^r = \bar{P}^{-1}(S_i^b) \subset L_q^r$, then

$$\bar{P}(\bar{S}_i^l) = S_i^t \subset \Pi_p \quad \text{and} \quad \bar{P}(\bar{S}_i^r) = S_i^b \subset \Pi_p$$

It follows from choice of $\Pi_\varsigma(\bar{\varsigma}_n, \varsigma_n) \subset \Pi_q$ and **Fact 2** that

$$0 < z_p < \bar{\varepsilon}_p \text{ for every point } (\phi, z_p) \in \bar{P}(\bar{S}_i^l \cup \bar{S}_i^r).$$

Therefore for every point $(\bar{\theta}, \bar{z}_q) \in P(\bar{S}_i^l \cup \bar{S}_i^r) = P_4 \circ (\bar{P}(\bar{S}_i^l \cup \bar{S}_i^r))$ there holds

$$|\bar{\theta} - \theta_q| < \eta. \tag{23}$$

On the other hand, since $\bar{S}_i^t = \bar{P}^{-1}(S_i^l)$ and $\bar{S}_i^b = \bar{P}^{-1}(S_i^r)$, $i = 1,2,\ldots,n$, we have

$$\bar{P}(\bar{S}_i^t) = S_i^l \subset \Pi_p \quad \text{and} \quad \bar{P}(\bar{S}_i^b) = S_i^r \subset \Pi_p$$

And in view of **Fact 2,** we have

$$\bar{z}_q \geq \bar{\varepsilon}_q \quad \text{for each point } (\bar{\theta}, \bar{z}_q) \in P(\bar{S}_i^t) = P_4 \circ (\bar{P}(\bar{S}_i^t)) \tag{24}$$

and

$$\bar{z}_q < 0 \quad \text{for } (\bar{\theta}, \bar{z}_q) \in P(\bar{S}_i^r) = P_4 \circ (\bar{P}(\bar{S}_i^r)) \tag{25}$$

if $g > 0$, or

$$\bar{z}_q \geq \bar{\varepsilon}_q \quad \text{for each point } (\bar{\theta}, \bar{z}_q) \in P(\bar{S}_i^r) = P_4 \circ (\bar{P}(\bar{S}_i^r)) \tag{26}$$

and

$$\bar{z}_q < 0 \quad \text{for } (\bar{\theta}, \bar{z}_q) \in P(\bar{S}_i^t) = P_4 \circ (\bar{P}(\bar{S}_i^t)). \tag{27}$$

if $g < 0$, as illustrated in Fig. 7 and Fig.8

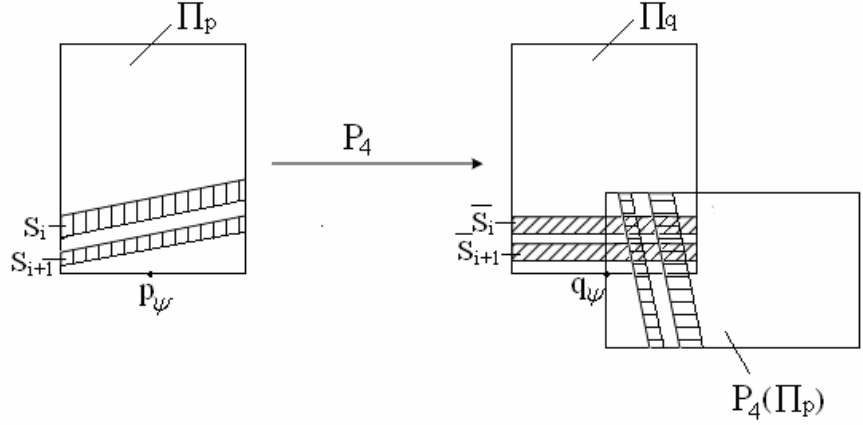

**Fig.7** The image $P_4 \circ (\overline{P}(\overline{S}_i^r))$ for $g > 0$

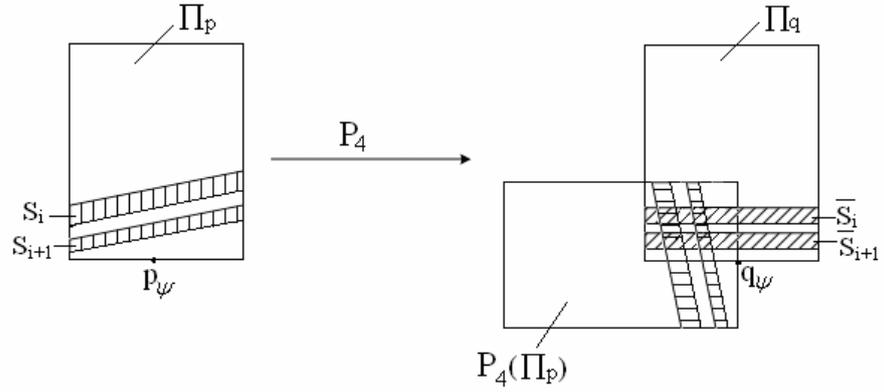

**Fig.8** The image $P_4 \circ (\overline{P}(\overline{S}_i^r))$ for $g < 0$

It can be seen from (23)-(27) that the disjoint compact subsets $\overline{S}_i \subset \Pi_\varsigma(\overline{\varsigma}_n, \varsigma_n)$, $i = 1, 2, ..., n$, are the candidates for existence of horseshoe $\Lambda_n$ of $P$ in $\Pi_\varsigma(\overline{\varsigma}_n, \varsigma_n)$, and the proof is thus completed. □

### 3 Further discussions

It is obvious that the assumption **A1** can be relaxed as follows

At the point $p$, the linearized flow of (1) in an appropriate local coordinate system is given by the following equations

$$\begin{cases} \dot{x}_p = \lambda_1 x_p \\ \dot{y}_p = \lambda_2 y_p \qquad \lambda_{1,2} > 0 \text{ and } \tilde{\lambda} > 0 \\ \dot{z}_p = -\tilde{\lambda} z_p \end{cases} \qquad (28)$$

In this case the cross section $\Pi_1$ in the neighborhood of $p$ can be chosen as shown in Fig.?

$$\Pi_1 = \{(x_p, y_p, z_p) : |x_p| \leq \nu, y_p = \nu, 0 < z_p < \delta_p\}$$

And replace the section $\pi_2 = \{(x_p, y_p, z_p) : x_p^2 + y_p^2 < r_p^2, z_p = \delta_p\}$ with

$$\Pi_2 = \{(x_p, y_p, z_p) : |x_p| \leq \nu, |y_p| \leq \nu, z_p = \delta_p\}$$

In view of (4) the map $\hat{P} : \Pi_1 \to \Pi_2$ can be defined as follows.

$$\hat{P}(x_p, \nu, z_p) = \begin{pmatrix} x(\dfrac{z_p}{\nu})^{\lambda_1/\tilde{\lambda}} \\ \nu(\dfrac{z_p}{\nu})^{\lambda_2/\tilde{\lambda}} \\ \nu \end{pmatrix} \qquad (29)$$

In the arguments the map $P_1 : \hat{\Pi}_2 \subset \Pi_2 \to \Pi_1$ is taken to be $P_1 = \hat{P}^{-1}$, where

$$\hat{\Pi}_2 = \hat{P}(\Pi_1)$$

As for the composition $P : \Pi_q \to \Pi_q$

$$P = P_4 \circ P_1 \circ P_2 \circ P_3$$

The arguments are similar to that of proof of **Theorem 1**, after we talking something about the map $P_1 = \hat{P}^{-1}$. Without loss of generality, the **Remark 2** can be replaced by the assumption that the point $(x_p, y_p, z_p) = (0, \nu, 0)$ is the point at which the heteroclinic orbit $\psi(t)$ intersects with $\Pi_1$.

The image of the map $\hat{P} : \Pi_1 \to \Pi_2$ is something shaped like a wedge in $\Pi_2$

It is easy to see that $\hat{P}(\Pi_1) \bigcap P_2 \circ P_3(\Pi_q)$ consists of a number of disjoint connected components, and that $P_1 \circ \hat{P}(\Pi_1) \bigcap P_2 \circ P_3(\Pi_q)$ also consist of the same number of disjoint connected components. Then for a properly chosen $\Pi_\varsigma(\bar{\varsigma}_n, \varsigma_n))$ consider the disjoint connected components of $\Pi_p \bigcap P_1 \circ P_2 \circ P_3(\Pi_\varsigma(\bar{\varsigma}_n, \varsigma_n))$ as in the proof of **Theorem 1.** □

The following statement is also obvious.

**Theorem 2** Let $X^r$ be the space of $C^r$ vector fields on $R^n$. Suppose that $F(x)$ be the vector field satisfying the assumptions **A1-A4**. Then there is a neighborhood $N$ of $F(x)$, such that for each vector field $Y(x) \in N$, there is some integer $m > 1$ such that there exists an invariant compact set in $\Lambda_m \subset \Pi_q$ such that the Poincaré map $P: \Pi_q \to \Pi_q$ restricted to $\Lambda_m$ is semi-conjugate to $m$-shift map.

*Remark*   After the completion of this paper we find that V. V. Bykov obtained independently a result similar to **Theorem 1** with a brief argument that seems hard to catch   (V. V. Bykov , The bifurcations of separatrix contours and chaos,   Physica D: Nonlinear Phenomena, Volume 62, Issues 1-4, 30 January 1993, Pages 290-299). Our arguments for are rigorous and more geometrically transparent

**Acknowledgements** This work is supported in part by National Natural Science Foundation of China (10972082).    This work is also partial supported by "Program for Visiting Scholars at the Chern Institute of Mathematics"